\RequirePackage{fix-cm} 
\documentclass[smallextended]{svjour3}     
\smartqed  
\usepackage{graphicx}
 \journalname{Hyperfine Interactions 
}
\begin{document}

\title{Consistency of $\Lambda\Lambda$ hypernuclear events}
\titlerunning{$\Lambda\Lambda$ hypernuclei} 

\author{A. Gal \and D.J. Millener}

\authorrunning{A.~Gal,~D.J. Millener} 

\institute{A. Gal \at
              Racah Institute of Physics, The Hebrew University,
Jerusalem~91904, Israel \\ 
\and D.J. Millener \at Physics Department, Brookhaven National Laboratory, 
Upton, NY 11973, USA \\  
} 

\date{Received: date / Accepted: date}

\maketitle

\begin{abstract}

Highlights of $\Lambda\Lambda$ emulsion events are briefly reviewed. 
Given three accepted events, shell-model predictions based on 
$p$-shell $\Lambda$ hypernuclear spectroscopic studies are shown to 
reproduce $B_{\Lambda\Lambda}({_{\Lambda\Lambda}^{~10}{\rm Be}})$ and 
$B_{\Lambda\Lambda}({_{\Lambda\Lambda}^{~13}{\rm B}})$ in terms of 
$B_{\Lambda\Lambda}({_{\Lambda\Lambda}^{~~6}{\rm He}})$. Predictions for 
other species offer judgement on several alternative assignments of the 
$_{\Lambda\Lambda}^{~13}{\rm B}$ KEK-E176 event, and on the assignments 
$_{\Lambda\Lambda}^{~11}{\rm Be}$ and $_{\Lambda\Lambda}^{~12}{\rm Be}$ 
suggested recently for the KEK-E373 HIDA event. 
The predictions of the shell model, spanning a wide range of $A$ values, 
are compared with those of cluster models, where the latter are available. 


\end{abstract}

\section{Introduction}
\label{sec:intro}

$\Lambda\Lambda$ hypernuclei provide valuable information on the 
$\Lambda\Lambda$ interaction and how it fits into our understanding 
of the baryon-baryon interaction. Although the existence of 
$\Lambda\Lambda$ hypernuclei nearly rules out a stable $H$ dibaryon, 
a $\Xi N$ dominated $H$ resonance might affect the systematics of 
$\Lambda\Lambda$ binding energies. Only three emulsion events presented 
serious candidates for $\Lambda\Lambda$ hypernuclei before 2001: 
$_{\Lambda\Lambda}^{~10}$Be \cite{Dan63,DDF89}, $_{\Lambda\Lambda}^{~~6}$He 
\cite{Pro66} and $_{\Lambda\Lambda}^{~13}$B \cite{Aok91,DMG91}. 
The $\Lambda\Lambda$ binding energies $B_{\Lambda\Lambda}$ deduced from 
these events indicated that the ${^{1}S_0}$ interaction $V_{\Lambda\Lambda}$ 
was strongly attractive, with a $\Lambda\Lambda$ excess binding energy 
$\Delta B_{\Lambda\Lambda} \sim 4.5$ MeV, although it had been 
realized that the binding energies of $_{\Lambda\Lambda}^{~10}$Be and 
$_{\Lambda\Lambda}^{~~6}$He were inconsistent with each other \cite{BUC84}. 
Here, the $\Lambda\Lambda$ excess binding energy is defined as 
\begin{equation} 
\label{eq:delB} 
\Delta B_{\Lambda\Lambda} (^{~~A}_{\Lambda \Lambda}Z) 
= B_{\Lambda\Lambda} (^{~~A}_{\Lambda \Lambda}Z) 
- 2{\bar B}_{\Lambda} (^{(A-1)}_{~~~~\Lambda}Z)\;, 
\end{equation} 
where 
${\bar B}_{\Lambda}$ is the (2$J$+1)-average of 
$B_{\Lambda}$ values for the $^{(A-1)}_{~~~~\Lambda}Z$ hypernuclear core 
levels. For comparison, $\Delta B_{\Lambda N}({_{\Lambda}^{5}{\rm He}})=
1.73\pm 0.13$ MeV, implying the unnatural ordering $\Delta B_{\Lambda\Lambda} 
> \Delta B_{\Lambda N}$. This perception changed in 2001 when a uniquely 
assigned $_{\Lambda\Lambda}^{~~6}{\rm He}$ hybrid-emulsion event \cite{Tak01}, 
with updated values \cite{Nak10}  
\begin{equation} 
\label{eq:LL6He} 
B_{\Lambda\Lambda}(_{\Lambda\Lambda}^{~~6}{\rm He})=6.91\pm 0.16~{\rm MeV}, 
\;\;\;\; \Delta B_{\Lambda\Lambda}(_{\Lambda\Lambda}^{~~6}{\rm He})=0.67\pm 
0.17~{\rm MeV} \;, 
\end{equation} 
ruled out the high value of $\Delta B_{\Lambda\Lambda}$ from the dubious 
earlier $_{\Lambda\Lambda}^{~~6}{\rm He}$ event \cite{Pro66}, restoring thus 
the expected hierarchy $\Delta B_{\Lambda\Lambda} < \Delta B_{\Lambda N}$. 
Both capture at rest formation $\Xi^- +{^{12}{\rm C}}\to{_{\Lambda\Lambda}^
{~~6}{\rm He}} + t + \alpha$ and weak decay $_{\Lambda\Lambda}^{~~6}
{\rm He}\to {_{\Lambda}^{5}{\rm He}}+p+\pi^-$, in this so called NAGARA event, 
yield consistently with each other the values listed in (\ref{eq:LL6He}). 
Neither $_{\Lambda\Lambda}^{~~6}{\rm He}$ nor $_{\Lambda}^{5}{\rm He}$ 
have excited states that could bias the determination of $B_{\Lambda\Lambda}
(_{\Lambda\Lambda}^{~~6}{\rm He})$. 


Accepting the NAGARA event calibration of $V_{\Lambda\Lambda}$, we review 
and discuss (i) particle stability for lighter $\Lambda\Lambda$ hypernuclei; 
(ii) reinterpretation of the events assigned $_{\Lambda\Lambda}^{~10}{\rm Be}$ 
and $_{\Lambda\Lambda}^{~13}{\rm B}$; (iii) several alternative assignments 
for the $_{\Lambda\Lambda}^{~13}{\rm B}$ event; 
and (iv) plausibility of the assignments $_{\Lambda\Lambda}^{~11}{\rm Be}$ 
or $_{\Lambda\Lambda}^{~12}{\rm Be}$ proposed for the recently reported HIDA 
event \cite{Nak10}. In the course of doing so, we compare $B_{\Lambda\Lambda}$ 
values derived from emulsion events with shell-model predictions \cite{GMi11} 
and with selected few-body cluster calculations \cite{BUC84,HKM02,HKY10} where 
the latter exist.

\section{Onset of $\Lambda\Lambda$ hypernuclear stability} 
\label{sec:onset} 

From the very beginning it was recognized that $\Lambda\Lambda$ and 
$\Lambda\Lambda N$ were unbound \cite{Dal63b,THe65}; if $\Lambda\Lambda N$ 
were bound, the existence of a $nn\Lambda$ bound state would follow. 
The existence of a $_{\Lambda\Lambda}^{~~4}$H bound state was claimed by 
AGS-E906 \cite{Ahn01a}, from correlated weak-decay pions emitted sequentially 
by $\Lambda\Lambda$ hypernuclei produced in a $(K^-,K^+)$ reaction on $^9$Be. 
However, the $_{\Lambda\Lambda}^{~~4}$H interpretation is controversial 
\cite{KFO02,RHu07}. Several post-2001 calculations exist for 
$^{~~4}_{\Lambda\Lambda}$H. A Faddeev-Yakubovsky 4-body calculation finds 
no bound state \cite{FGa02c}, whereas a stochastic-variational (SV) 4-body 
calculation finds it to be bound by as much as 0.4 MeV \cite{NAM03}. 
The more comprehensive $s$-shell $\Lambda$- and $\Lambda\Lambda$-hypernuclear 
SV calculation of Ref.~\cite{NSA05} finds $^{~~4}_{\Lambda\Lambda}$H to be 
particle stable by as little as a few keV, which would be insufficient to 
maintain particle stability once $V_{\Lambda\Lambda}$ is renormalized to 
reproduce the recently updated (smaller) value of Eq.~(\ref{eq:LL6He}) for 
$\Delta B_{\Lambda\Lambda}(_{\Lambda\Lambda}^{~~6}{\rm He})$. 

\begin{figure*}[htb] 
\includegraphics[width=0.7\textwidth]{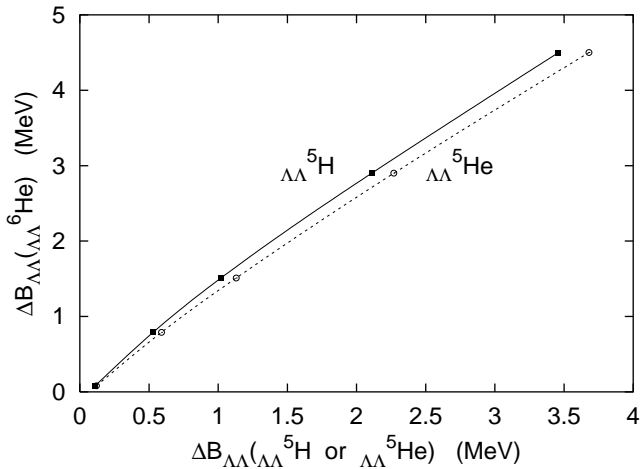} 
\caption{Faddeev calculations of 
$\Delta B_{\Lambda\Lambda}({_{\Lambda\Lambda}^{~~6}{\rm He}})$ {\it vs.} 
$\Delta B_{\Lambda\Lambda}({_{\Lambda\Lambda}^{~~5}{\rm H}},
{_{\Lambda\Lambda}^{~~5}{\rm He}})$ \cite{FGa02b}, see text.} 
\label{fig:2}
\end{figure*} 

Regardless of whether $^{~~4}_{\Lambda\Lambda}$H is particle-stable or not, 
there is a general consensus that the mirror $\Lambda\Lambda$ hypernuclei 
$^{~~5}_{\Lambda\Lambda}$H--$^{~~5}_{\Lambda\Lambda}$He are particle-stable, 
with $\Delta B_{\Lambda\Lambda}\sim 0.5-1$ MeV \cite{FGa02b,FGS03}, or larger 
owing to the $\Lambda\Lambda-\Xi N$ coupling which is particularly effective 
here \cite{NSA05,MSA03,LYa04}. In addition, substantial charge symmetry 
breaking effects are expected in these systems, resulting in a higher 
binding energy of $^{~~5}_{\Lambda\Lambda}$He by up to 0.5 MeV with respect 
to $^{~~5}_{\Lambda\Lambda}$H \cite{LYa04,YRi08}. Figure~\ref{fig:2} 
demonstrates how $\Delta B_{\Lambda\Lambda}$ values for the $A=5,6$ systems, 
calculated over a broad range of $V_{\Lambda\Lambda}$ strengths, are nearly 
linearly correlated with only a small offset. Thus, the stability of 
$^{~~6}_{\Lambda\Lambda}$He ensures stability for $^{~~5}_{\Lambda\Lambda}$H.

\section{Ingredients of hypernuclear shell model}
\label{sec:SM} 

Shell-model predictions for $\Lambda\Lambda$ hypernuclei have been 
given recently \cite{GMi11} using Eq.~(\ref{eq:delB}) in which 
$\Delta B_{\Lambda\Lambda}(^{~~A}_{\Lambda \Lambda}Z)$ is replaced by 
a constant $V_{\Lambda\Lambda}$ matrix element, identified with 
$\Delta B_{\Lambda\Lambda}(_{\Lambda\Lambda}^{~~6}{\rm He})$ of 
Eq.~(\ref{eq:LL6He}).{\footnote{A straightforward modification for 
$_{\Lambda\Lambda}^{~10}{\rm Be}_{\rm g.s.}(0^+)$, with a nuclear core 
$^8$Be unstable to $\alpha$ emission, is discussed in Ref.~\cite{GMi11}.}} 
Calculations of $B_{\Lambda\Lambda}(^{~~A}_{\Lambda\Lambda}Z)$ require then 
the knowledge of ${\bar B}_{\Lambda}(^{(A-1)}_{~~~~\Lambda}Z)$ involving 
single-$\Lambda$ hypernuclear ground-state (g.s.) binding energies plus 
g.s. doublet splittings $\Delta E_{\rm g.s.}$ for $J_{\rm core}\neq 0$. 
Table~\ref{tab:1} lists $\Delta E_{\rm g.s.}$ values relevant for the 
calculations reviewed here, exhibiting remarkable agreement between 
theory and experiment. 

\begin{table}[htb] 
\caption{Doublet splittings $\Delta E^{\rm th}$ and $\Delta E^{\rm exp}$ 
(in keV) from Refs.~\cite{Mil10,Tam05}, where $\Delta E^{\rm th}_{\rm alt}$ 
uses ESC04a--inspired $\Lambda-\Sigma$ coupling. Note the sensitivity 
of $\Delta E^{\rm th}({_{~\Lambda}^{10}{\rm B}_{\rm g.s.}})$ to the model 
used for $\Lambda-\Sigma$ mixing. The $_{\Lambda}^{9}{\rm Be}^{\ast}$ 
and $_{~\Lambda}^{13}{\rm C}^{\ast}$ excited doublets are discussed in 
Sect.~\ref{sec:consistency}.}  
\label{tab:1} 
\begin{tabular}{lccccc} 
\hline\noalign{\smallskip} 
& $J_{\rm up}^{\pi}$ & $J_{\rm low}^{\pi}$ & $\Delta E^{\rm th}$ & 
$\Delta E^{\rm th}_{\rm alt}$ & $\Delta E^{\rm exp}$ \\ 
\noalign{\smallskip}\hline\noalign{\smallskip} 
$_{\Lambda}^{9}{\rm Be}^{\ast}$ & $3/2^+$ & $5/2^+$ & 44 & 49 & 43$\pm$5 \\ 
\noalign{\smallskip} 
$_{~\Lambda}^{10}{\rm B}_{\rm g.s.}$ & $2^-$ & $1^-$ & 120 & 34 & 
$\leq 100$ \\ 
\noalign{\smallskip} 
$_{~\Lambda}^{11}{\rm B}_{\rm g.s.}$ & $7/2^+$ & $5/2^+$ & 267 & 243 & 
262.9$\pm$0.2 \\ 
\noalign{\smallskip} 
$_{~\Lambda}^{12}{\rm C}_{\rm g.s.}$ & $2^-$ & $1^-$ & 153 & 167 & 
161.4$\pm$0.7 \\ 
\noalign{\smallskip} 
$_{~\Lambda}^{13}{\rm C}^{\ast}$ & $5/2^+$ & $3/2^+$ & 31 & 47 & -- \\ 
\noalign{\smallskip}\hline 
\end{tabular} 
\end{table}

\section{Interpretation of $_{\Lambda\Lambda}^{~10}{\rm Be}$ 
and $_{\Lambda\Lambda}^{~13}{\rm B}$ emulsion events 
} 
\label{sec:consistency} 

The $B_{\Lambda\Lambda}$ values of both $_{\Lambda\Lambda}^{~10}{\rm Be}$ 
($17.5\pm 0.4$ MeV) \cite{DDF89} and $_{\Lambda\Lambda}^{~13}{\rm B}$ 
($28.2\pm 0.7$ MeV) \cite{Aok09} were extracted assuming that their $\pi^-$ 
weak decay proceeds to the g.s. of the respective daughter 
$\Lambda$ hypernuclei. This led to $\Delta B_{\Lambda\Lambda}\sim 4-5$ MeV, 
substantially higher than for $_{\Lambda\Lambda}^{~~6}{\rm He}$ (NAGARA). 
However, as realized by Danysz {\it et al.} \cite{Dan63}, the decay could 
proceed to excited states of the daughter $\Lambda$ hypernucleus which 
deexcites then rapidly to the g.s. emitting unobserved $\gamma$ radiation. 
This reduces the apparent $B_{\Lambda\Lambda}$ and $\Delta B_{\Lambda\Lambda}$ 
values by the $\Lambda$ hypernuclear excitation energy involved in the $\pi^-$ 
weak decay. Consistency with $_{\Lambda\Lambda}^{~~6}{\rm He}$ is restored 
upon accepting the following weak decays: 
\begin{equation} 
_{\Lambda\Lambda}^{~10}{\rm Be} \rightarrow {_{\Lambda}^{9}{\rm Be}}^{\ast}
(3/2^+,5/2^+;3.04~{\rm MeV}) + p + \pi^- , 
\label{eq:exc9} 
\end{equation} 
\begin{equation} 
_{\Lambda\Lambda}^{~13}{\rm B} \rightarrow {_{~\Lambda}^{13}{\rm C}}^{\ast}
(3/2^+,5/2^+;4.9~{\rm MeV}) + \pi^-, 
\label{eq:exc13} 
\end{equation} 
with rates comparable to those for decays to $_{\Lambda}^{9}{\rm Be}_{\rm g.s.}
(1/2^+)$ and $_{~\Lambda}^{13}{\rm C}_{\rm g.s.}(1/2^+)$, respectively. 
The doublet splittings of $_{\Lambda}^{9}{\rm Be}^{\ast}$ and 
$_{~\Lambda}^{13}{\rm C}^{\ast}$ are listed in Table~\ref{tab:1}. 

$_{\Lambda\Lambda}^{~10}{\rm Be}$ also fits the Demachi-Yanagi event 
observed in KEK-E373 \cite{Ahn01b}, with $B_{\Lambda\Lambda}=11.90\pm 0.13$ 
MeV \cite{Nak10} determined from the assumed formation reaction kinematics. 
The approximately 6 MeV difference between this and the 
Danysz {\it et al.} \cite{Dan63,DDF89} value for 
$B_{\Lambda\Lambda}({_{\Lambda\Lambda}^{~10}{\rm Be}})$ is reconciled by 
assuming that the Demachi-Yanagi event corresponds to formation of the 
first excited state $_{\Lambda\Lambda}^{~10}{\rm Be}^{\ast}$, 
\begin{equation} 
\Xi^- + {^{12}{\rm C}} \rightarrow 
{_{\Lambda\Lambda}^{~10}{\rm Be}^{\ast}(2^+;\approx 3~{\rm MeV})} + t, 
\label{eq:exc10} 
\end{equation} 
which decays to $_{\Lambda\Lambda}^{~10}{\rm Be}_{\rm g.s.}$ by emitting 
unseen $\gamma$ ray, the energy of which has to be added to the apparent 
$B_{\Lambda\Lambda}$ value deduced by assuming a g.s. formation. 
It is not clear why the formation of $_{\Lambda\Lambda}^{~10}{\rm Be}^{\ast}$ 
should be comparable or enhanced with respect to that of 
$_{\Lambda\Lambda}^{~10}{\rm Be}_{\rm g.s.}$. 

The $B_{\Lambda\Lambda}^{\rm exp}$ values corresponding to 
Eqs.~(\ref{eq:exc9})--(\ref{eq:exc10}) are listed in Table~\ref{tab:2} 
together with predictions made in cluster model (CM) and shell model (SM) 
calculations, all of which use $\Lambda\Lambda$ interactions normalized to 
$B_{\Lambda\Lambda}(_{\Lambda\Lambda}^{~~6}{\rm He})=6.91\pm 0.16$ MeV 
\cite{Nak10}. For $_{\Lambda\Lambda}^{~13}{\rm B}$, assuming charge symmetry, 
the $_{~\Lambda}^{12}{\rm B}_{\rm g.s.}$ doublet splitting input was 
identified with that of $_{~\Lambda}^{12}{\rm C}_{\rm g.s.}$ from 
Table~\ref{tab:1}. It is seen that both CM and SM calculations reproduce the 
reinterpreted $B_{\Lambda\Lambda}$ values of $_{\Lambda\Lambda}^{~10}{\rm Be}$ 
and $_{\Lambda\Lambda}^{~13}{\rm B}$. The SM agrees well with the 
Hiyama {\it et al.} CM calculation \cite{HKM02,HKY10}, and the SM 
calculation has no match for $_{\Lambda\Lambda}^{~13}{\rm B}$.  

\begin{table}[htb] 
\caption{Reinterpreted $B_{\Lambda\Lambda}^{\rm exp}$ values (in MeV) and 
predictions based on the NAGARA event for $_{\Lambda\Lambda}^{~~6}{\rm He}$. 
The error on $B_{\Lambda\Lambda}^{\rm exp}({_{\Lambda\Lambda}^{~~6}{\rm He}})$ 
is incorporated into the predicted values.} 
\label{tab:2} 
\begin{tabular}{cccccc} 
\hline\noalign{\smallskip}
& \multicolumn{2}{c}{$B_{\Lambda\Lambda}^{\rm exp}$} & 
\multicolumn{2}{c}{$B_{\Lambda\Lambda}^{\rm CM}$} & 
\multicolumn{1}{c}{$B_{\Lambda\Lambda}^{\rm SM}$} \\ 
& Eqs.~(\ref{eq:exc9},\ref{eq:exc13})& Eq.~(\ref{eq:exc10}) & 
\cite{BUC84} & \cite{HKY10} & \cite{GMi11} \\ 
\noalign{\smallskip}\hline\noalign{\smallskip}
$_{\Lambda\Lambda}^{~10}{\rm Be}$ & $14.5\pm 0.4$ & $14.94\pm 0.13$ & 
$14.35\pm 0.19$ & $14.74\pm 0.19$ & $14.97\pm 0.22$ \\ 
\noalign{\smallskip} 
$_{\Lambda\Lambda}^{~13}{\rm B}$ & $23.3\pm 0.7$ & &--&--& 
$23.21\pm 0.21$ \\ 
\noalign{\smallskip}\hline
\end{tabular}
\end{table}

\section{Alternative interpretations of the $_{\Lambda\Lambda}^{~13}{\rm B}$ 
event} 
\label{sec:L13B} 

The emulsion event assigned to $_{\Lambda\Lambda}^{~13}{\rm B}$ 
\cite{Aok91,DMG91} has been carefully scrutinized by the KEK-E176 
Collaboration \cite{Aok09}. Several alternative assignments were pointed out, 
two of which that do not require $\Lambda$ hypernuclear excitation in the 
$\pi^-$ weak decay of the $\Lambda\Lambda$ hypernuclear g.s. are listed in 
Table~\ref{tab:3}. Comparison with model calculations suggests that such 
reassignments cannot be ruled out, although a $_{\Lambda\Lambda}^{~13}{\rm B}$ 
assignment shows a higher degree of consistency between the 
$B_{\Lambda\Lambda}$ values derived from formation and from decay. 
In particular, the accepted formation reaction $\Xi^- + {^{14}{\rm N}} \to 
{_{\Lambda\Lambda}^{~13}{\rm B}} + p + n$ was shown to occur naturally in 
$\Xi^-$ capture at rest in light nuclei emulsion \cite{DMG91}.  

\begin{table}[htb] 
\caption{Reassignments of the $_{\Lambda\Lambda}^{~13}{\rm B}$ 
KEK-E176 event. $B_{\Lambda\Lambda}$ values are in MeV.} 
\label{tab:3} 
\begin{tabular}{cccc} 
\hline\noalign{\smallskip}
& $B_{\Lambda\Lambda}^{\rm exp}$ \cite{Aok09} & $B_{\Lambda\Lambda}^{\rm CM}$ 
\cite{HKY10} & $B_{\Lambda\Lambda}^{\rm SM}$ \cite{GMi11} \\ 
\noalign{\smallskip}\hline\noalign{\smallskip}
$_{\Lambda\Lambda}^{~11}{\rm Be}$ & $17.53\pm 0.71$ & $18.23\pm 0.19$ & 
$18.40\pm 0.28$ \\ 
\noalign{\smallskip} 
$_{\Lambda\Lambda}^{~12}{\rm B}$ & $20.60\pm 0.74$ & -- & $20.85\pm 0.20$ \\ 
\noalign{\smallskip}\hline
\end{tabular}
\end{table}

\section{Interpretation of the KEK-E373 HIDA event} 
\label{sec:HIDA}

\begin{table}[htb] 
\caption{Assignments suggested for the KEK-E373 HIDA event. 
$B_{\Lambda\Lambda}$ values are in MeV.} 
\label{tab:4} 
\begin{tabular}{cccc} 
\hline\noalign{\smallskip}
& $B_{\Lambda\Lambda}^{\rm exp}$ \cite{Nak10} & $B_{\Lambda\Lambda}^{\rm CM}$ 
\cite{HKY10} & $B_{\Lambda\Lambda}^{\rm SM}$ \cite{GMi11} \\ 
\noalign{\smallskip}\hline\noalign{\smallskip}
$_{\Lambda\Lambda}^{~11}{\rm Be}$ & $20.83\pm 1.27$ & $18.23\pm 0.19$ & 
$18.40\pm 0.28$ \\ 
\noalign{\smallskip} 
$_{\Lambda\Lambda}^{~12}{\rm Be}$ & $22.48\pm 1.21$ & -- & $20.72\pm 0.20$ \\ 
\noalign{\smallskip}\hline 
\end{tabular} 
\end{table} 

The KEK-E373 Collaboration has recently presented evidence from the HIDA event 
for another $\Lambda\Lambda$ hypernucleus, tentatively assigned to either 
$_{\Lambda\Lambda}^{~11}{\rm Be}$ or to $_{\Lambda\Lambda}^{~12}{\rm Be}$ 
\cite{Nak10}. The associated $B_{\Lambda\Lambda}^{\rm exp}$ values, together 
with model predictions, are listed in Table~\ref{tab:4}. We note that since no 
experimental data exist on $_{\Lambda}^{11}{\rm Be}$, the required input for 
evaluating $B_{\Lambda\Lambda}^{\rm SM}({_{\Lambda\Lambda}^{~12}{\rm Be}})$ 
was derived within the SM approach \cite{GMi11}. It is clear from the table 
that neither of the proposed assignments is favorable, although the relatively 
large experimental uncertainties do not completely rule out either of these.

\section{Conclusion} 
\label{sec:concl} 

It was shown how the three acceptable $\Lambda\Lambda$ emulsion events, 
corresponding to $_{\Lambda\Lambda}^{~~6}{\rm He}$, 
$_{\Lambda\Lambda}^{~10}{\rm Be}$ and $_{\Lambda\Lambda}^{~13}{\rm B}$, 
can be made consistent with each other, in good agreement with CM and with 
SM calculations of $B_{\Lambda\Lambda}$. Other possible assignments for 
the KEK-E176 $_{\Lambda\Lambda}^{~13}{\rm B}$ event were discussed, and 
the assignments proposed for the recently reported HIDA event were found 
unfavorable. It was pointed out that simple shell-model estimates, making 
use of $\Lambda$-hypernuclear spectroscopic data and analysis, are sufficient 
for discussing the world data of $\Lambda\Lambda$ hypernuclear events. 
A relatively weak $\Lambda\Lambda$ interaction, with $(1s_{\Lambda})^2$ 
matrix element of magnitude $\Delta B_{\Lambda\Lambda}(_{\Lambda\Lambda}^{~~6}
{\rm He})=0.67\pm 0.17$ MeV, describes well the data in the observationally 
accessible mass range $6 \leq A \leq 13$. Comparably weak $\Lambda\Lambda$ 
interactions are obtained also in recent theoretical models, in Nijmegen 
extended soft-core (ESC) models \cite{YRi08,RNY10} and in lowest order 
$\chi$EFT \cite{PHM07}. Less well determined is the $\Lambda\Lambda$ coupling 
to the slightly higher $\Xi N$ channel, with appreciable model dependence in 
ESC models \cite{YRi08,RNY10}. The observation of $A=5$ $\Lambda\Lambda$ 
hypernuclei would add valuable new information on this issue.

\begin{acknowledgements}
Useful discussions with Emiko Hiyama on hypernuclear cluster-model 
calculations are gratefully acknowledged. 
\end{acknowledgements}

\end{document}